\documentclass[twocolumn,showpacs,superscriptaddress,nofootinbib]{revtex4-1}
\usepackage[english]{babel}
\usepackage[pdftex]{graphicx}
\usepackage[colorlinks=true,citecolor=blue,urlcolor=blue,linkcolor=blue]{hyperref}
\usepackage{amsmath,amssymb}
\usepackage{units}
\usepackage{mathpazo}

\DeclareMathOperator*{\argmin}{arg\,min}

\begin{document}

\title{Maximum-likelihood reconstruction of photon returns from 
simultaneous analog and photon-counting lidar measurements}

\author{Darko Veberi\v{c}}
\email{\tt darko.veberic@ung.si}
\affiliation{Laboratory for Astropaticle Physics, University of Nova 
Gorica, Vipavska 13, SI-5000 Nova Gorica, Slovenia}
\affiliation{Department for Theoretical Physics, J.\ Stefan Institute, 
Jamova 39, SI-1000 Ljubljana, Slovenia}

\begin{abstract}
We present a novel method for combining the analog and photon-counting
measurements of lidar transient recorders into reconstructed photon 
returns. The method takes into account the statistical properties of the 
two measurement modes and estimates the most likely number of arriving 
photons and the most likely values of acquisition parameters describing 
the two measurement modes. It extends and improves the standard 
combining (``gluing'') methods and does not rely on any \emph{ad hoc} 
definitions of the overlap region nor on any background subtraction 
methods.
\end{abstract}


\maketitle

\section{Introduction}

In order to extend the total dynamic range of measurements the same 
back-scattered return signal in modern lidar acquisition systems (a.k.a.\ 
transient recorders) is usually sampled with two distinct methods: a 
fast analog-to-digital converter and a photon counting unit 
\cite{eichinger}. The two range-resolved traces are then combined 
(``glued'') by first correcting the dead-time effects of the photon 
counting mode \cite{donovan} and then by calibrating (fitting) the 
analog trace to the photon counts in some suitable photon-counting rate 
interval \cite{gluing,newsom,whiteman}.  The final step in the 
construction of the ``glued'' trace involves choosing a suitable signal 
size above which only rescaled analog values are considered and below 
which only the photon-counting trace is used.  The general usability of 
such ``gluing'' methods is hampered by several intrinsic weaknesses: the 
``background'' is usually subtracted from both measurement modes whereas 
it could be retained and used as additional information; a large variety 
of regressions and $\chi^2$ minimizations are used in the calibration of 
the analog signal; arbitrary and not well defined photon-counting rate 
fitting intervals are imposed in order to stabilize the former 
minimizations; photon-counting nonlinearity is usually corrected only 
with manufacturer-supplied dead-time values \cite{licel-gluing}; the 
unused half of the measured data is simply discarded and the actual 
position of the crossover between the analog and the photon-counting 
trace is not selected by any strict rules.

\section{Measurement}

This new method is based on the elementary observation that the 
\emph{same} input signal is evaluated by \emph{two different} 
measurement techniques. Employing simple models of these two measurement 
processes we will make use of \emph{all} available data to construct a 
\emph{new composite} trace of arriving photon numbers in such a way that 
the new hypothetical number of photons would in fact most likely produce 
the two actually measured traces.

For the sake of clarity we will keep the measurement models simple 
enough to illustrate the main properties of the method. Nevertheless, 
the procedure is highly flexible and if greater levels of detail are 
required, then the descriptions in Eq.~\eqref{analog}--\eqref{dtc-var} 
can be simply updated with more complex descriptions of the two 
measurement processes.

\subsection{Analog signal}

In a typical transient recorder, the analog signal is constructed by 
integrating the current from a photomultiplier (PMT) in a
sampling time $\Delta t$ which is then discretized by an 
analog-to-digital converter into the analog lidar trace. Since the PMT 
is a fairly linear sensor of arrived photons $p$, we can thus
describe their transformation into an analog signal $a$ with a simple 
linear transformation
\begin{equation}
a = A(p) = \alpha p + \beta,
\label{analog}
\end{equation}
where $\alpha$ is related to the PMT and amplifier gain, converting the 
number of incoming photons into the adc units. $\beta$ is a small 
hardware-imposed offset (baseline) which enables the detection of a 
possible signal undershoot and post-measurement determination of a true 
zero.  Given the number of the input photons $p$, the variance, $V[a]$, 
of the analog signal resulting from the noise in this chain of 
electronics can be, at least for small signals, safely modeled as being 
constant,
\begin{equation}
V[a] = \gamma^2
\label{analog-var}
\end{equation}
and is expressed in units of $\unit{adc^2}$. For larger signals the 
analog variance in Eq.~\eqref{analog-var} acquires an additional 
signal-size dependent Poisson term which we can neglect for reasons 
given later.

\subsection{Photon-counting mode}

In typical photon-counting modules of modern transient recorders, the 
input photons $p$ are recorded by counters with predominantly 
\emph{non-extending} dead-time $\tau$. These types of counters are also 
referred to as \emph{cumulative} or \emph{non-paralyzable} counters.

For such counters\footnote{The procedure given here can be naturally 
adapted also for the \emph{extending} (or \emph{paralyzable}) type of 
photon counters by replacing Eq.~\eqref{dtc-mean} with 
$C(p)=p\exp(\delta p)$ and its associated variance \cite{donovan}.} the 
mean number of counts $m$ in a sampling time $\Delta t$ can be expressed 
as
\begin{equation}
m = C(p) = \frac{p}{1+\delta p},
\label{dtc-mean}
\end{equation}
where $\delta$ is the fraction of dead-time vs.\ sampling time,
$\delta=\tau/\Delta t$. The variance of the photon count is
\begin{equation}
V[m] = V_\delta(p),
\label{dtc-var}
\end{equation}
where $V_\delta$ is a nontrivial function for the variance of the 
dead-time counter and is explained in greater detail in Appendix 
\ref{a:dtc}.

Note that the dead time, during which the counter is unable to record 
any incoming photons, induces saturation of the maximally possible 
counts to $m_\text{max}\approx\lim_{p\to\infty}C(p)=1/\delta$.  As 
mentioned before, the standard gluing procedure involves, before fitting 
the analog and photon counting traces, correcting the counts $m$ for the 
dead-time effects with the inverse of the function in 
Eq.~\eqref{dtc-mean},
\begin{equation}
p = C^{-1}(m) = \frac{m}{1-\delta m}.
\end{equation}
Unfortunately, the inverse has a singularity at $m=m_\text{max}$ and 
produces negative photon estimates $p$ for $m>m_\text{max}$. In the 
standard gluing procedure this, and the fact that the dead time $\tau$ 
is not well known, limits the range of useful data of the measured 
photon counting traces.

As we will see later on, the procedure developed here does not suffer 
from this drawbacks since only the non-singular function $C(p)$ is used 
and the estimation of the dead-time value is part of the method.

\subsection{Overlap region}

In the case of a large number of incoming photons, $p\gg1/\delta$, only 
the analog signal carries useful information due to the inevitable 
saturation of the photon counter. For a small photon influx the 
situation is reversed since the analog signal has reached the levels of 
the electronic noise while the photon-counting is in the ideal 
proportional mode with almost no dead-time effects. Therefore, outside 
of the relative overlap region, the quality of data of one or the other 
mode prevails. From the simple measurement models given above we thus 
require good accuracy in the overlap region and that the winning
model is supplying the correct solution far away from the overlap.

Due to these relaxed requirements we can approximate the variance of the 
dead-time affected photon counter in Eq.~\eqref{dtc-var} with the 
Poisson variance of the photon-count itself (see Appendix \ref{a:dtc}),
\begin{equation}
V[m] = m = C(p).
\label{dtc-var-simple}
\end{equation}

\subsection{Summation of lidar traces}

It is quite common practice in lidar measurements to additionally 
increase the dynamical range of the data acquisition by summation 
(\emph{time stacking}) of consecutive lidar returns. With fast 
laser-pulse repetition rates it is reasonable to assume that within the 
summation time the atmosphere does not introduce substantial sources of 
additional variance beyond the natural Poisson-like fluctuations of the 
backscattered photons.

Denoting by $a_\text{s}$ the sum of $N_\text{s}$ analog measurements $a$ 
at the same range of consecutive lidar traces and with $p_\text{s}$ the 
sum of the arrived photons, the photon conversion in Eq.~\eqref{analog} 
is transformed into $a_\text{s} = N_\text{s}A(p_\text{s}/N_\text{s}) = 
\alpha p_\text{s}+N_\text{s}\beta$. The variance in 
Eq.~\eqref{analog-var} scales as $V[a_\text{s}] = N_\text{s}\gamma^2$.

The mean photon count obtained by summation of $N_\text{s}$ consecutive 
lidar returns, $N_\text{s}C(p_\text{s}/N_\text{s})$, has a nice property 
of retaining the general form of Eq.~\eqref{dtc-mean} with the dead-time 
fraction effectively transformed
into $\delta/N_\text{s}$. Nevertheless, for large photon numbers $p$ the 
variance depends on summation in a non-linear way and has to be 
evaluated as $N_\text{s}V_\delta(p_\text{s}/N_\text{s})$.

The measurement models given above are thus, at least to some degree, 
invariant with respect to the summation as long as the following 
transformation of the acquisition parameters is taken into account,
\begin{equation}
\alpha\mapsto\alpha,\quad
\beta\mapsto\beta/N_\text{s},\quad
\gamma^2\mapsto \gamma^2/N_\text{s},\quad
\delta\mapsto N_\text{s}\delta.
\end{equation}

\section{Initial estimates}

\begin{figure}[t]
\centering
\includegraphics[width=\linewidth]{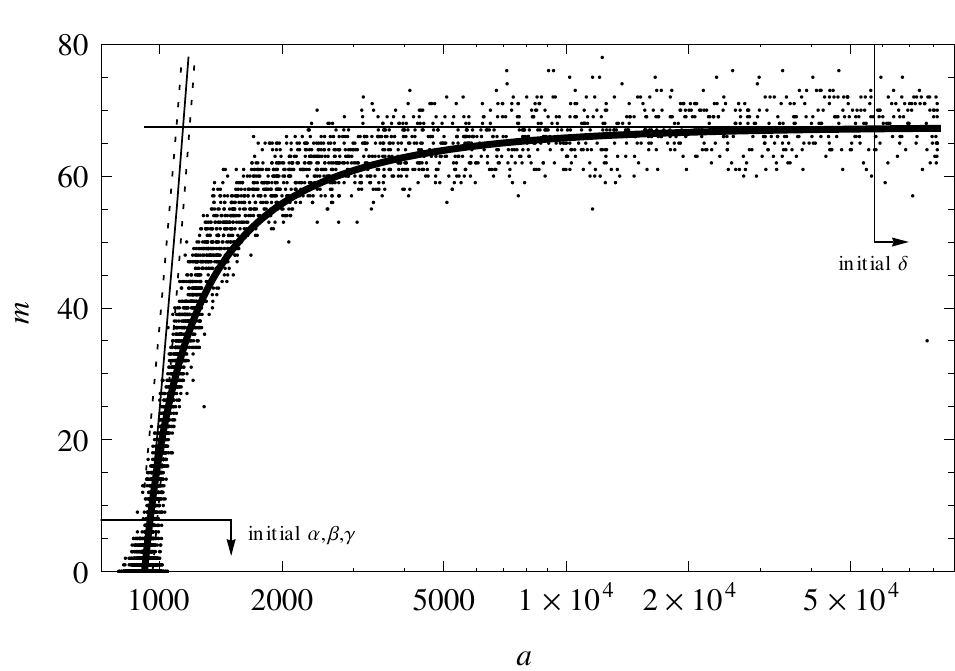}
\caption{Plot of analog vs.\ photon-counting data points $(a_i,m_i)$ for 
$N_\text{s}=20$ summed lidar returns.  The slanted line on the left is a 
fit for the photon-to-analog conversion parameters $\alpha$, $\beta$ and 
$\gamma^2$ in the range of the lower left corner (indicated by the left 
arrow-box) with dashed lines illustrating variance $a\pm2\gamma$.  The 
horizontal line is a fit for dead-time fraction $\delta$ in the range of 
the upper right corner (right arrow-box).  The thick line is the 
resulting prediction for $m=C(A^{-1}(a))$.}
\label{f:am}
\end{figure}

For any nonlinear minimization procedure it is of utmost importance to 
acquire accurate initial values for minimized parameters. In our case 
the initial values for parameters $\alpha$, $\beta$, $\gamma^2$ are 
obtained from a least-squares minimization of
\begin{equation}
\chi^2_\text{min} = \min_{\alpha,\beta} \sum_i [a_i - A(m_i)]^2,
\label{chi2}
\end{equation}
where only analog and photon-counting data points $(a_i,m_i)$ of the 
lower left corner are used (see Fig.~\ref{f:am}), i.e.\ the lower 10\% 
of the whole photon-count range.  Initial estimates for the ``gluing'' 
parameters $\alpha$ and $\beta$ are thus obtained in a manner similar to 
the standard procedure suggested by the manufacturers of the transient 
recorders \cite{licel-gluing} or other more detailed studies 
\cite{whiteman}.

Since the fluctuations of the analog data in the lower left corner of 
Fig.~\ref{f:am} are dominated by the electronic noise, an estimate for 
the analog variance $\gamma^2$ is obtained simply from the residuals in 
Eq.~\eqref{chi2}, $\gamma^2=\chi^2_\text{min}/N_\text{dof}$, where the 
number of degrees of freedom is $N_\text{dof}=N-2$ with $N$ the number 
of data points involved in the fit. From this point on, $\gamma^2$ is 
kept fixed and will not be the subject of minimization.

Fitting the photon counts $m$ to a constant in the tail of the large 
analog values (upper right corner of Fig.~\ref{f:am}) gives an estimate 
for the dead-time fraction, $\delta=1/\langle m\rangle$. For the 
$\langle m\rangle$ estimate typically only the data in the largest 30\% 
of the analog values has been used, but excluding all data points with 
ADC saturation (which we discard in this procedure anyway).

See Fig.~\ref{f:am} for the results of the initial fits to an example 
lidar return which will be used throughout this analysis. The data was 
obtained with a back-scatter lidar at wavelength of \unit[355]{nm}, 
pulse repetition rate of \unit[20]{Hz} and trace summation 
$N_\text{s}=20$.  The light sensor was a Hammamatsu R3200 
photomultiplier tube connected to a high-voltage of approximately 
\unit[800]{V}. The return signal was acquired with a Licel TR 40-160 
transient recorder operated at a sampling frequency of \unit[40]{MHz}.  
The recorder is delivering discretized analog signal traces with 
\unit[12]{bit} ADC resolution and photon-counting traces with a maximal 
count rate of \unit[250]{MHz}, both with trace depth of 16k samples 
\cite{licel}.  The example trace was recorded with $20^\circ$ elevation 
on a relatively clear night and contains only a thin and faint layer of 
haze around the range of \unit[13.5]{km}.

\section{Maximum likelihood}

From the two measurement models described above we can construct a 
likelihood $\mathcal{L}$ for the total trace as a product over all trace 
time elements $i$ of a likelihood of observing $p_i$ photons given the 
analog measurement $a_i$ and the photon count $m_i$,
\begin{equation}
\mathcal{L} =
  \prod_i\mathcal{L}(a_i,m_i,p_i),
\end{equation}
where likelihood $\mathcal{L}(a_i,m_i,p_i)$ is a product of the 
probability $P(a_i|p_i)$ to observe an analog signal and the probability 
to have a certain photon count $P(m_i|p_i)$ given a number of photons.  
We can model the analog probability with the normal (Gauss) distribution 
$\mathcal{N}(x,\sigma^2)=\exp(-x^2/2\sigma^2)/\sqrt{2\pi\sigma^2}$ using 
the linear transformation Eq.~\eqref{analog} and the corresponding 
variance $\gamma^2$. According to Eq.~\eqref{dtc-var-simple}, the photon 
count probability can be approximated with the Poisson distribution from
Eq.~\eqref{poisson} so that the resulting likelihood is expressed as
\begin{equation}
\mathcal{L}(a_i,m_i,p_i) =
  \mathcal{N}(a_i-A(p_i),\gamma^2)\times
  \mathcal{P}_{m_i}(C(p_i))
\label{likelihood}
\end{equation}
The corresponding deviance is defined as
\begin{align}
\mathcal{D} &=
  -2\ln\mathcal{L} =
  -2\sum_i\ln\mathcal{L}(a_i,m_i,p_i) =
\nonumber
\\
  &= \sum_i\mathcal{D}(a_i,m_i,p_i),
\label{deviance}
\end{align}
where
\begin{align}
\mathcal{D}(a_i,m_i,p_i) &=
  \ln2\pi\gamma^2 +
  \frac{[a_i-A(p_i)]^2}{\gamma^2} \,+
\nonumber
\\
  &+ 2\left[\ln m_i! + C(p_i) - m_i\ln C(p_i)\right]
\label{deviance1}
\end{align}
is the deviance for a particular data point\footnote{with specific 
requirement that $0\times\ln0\equiv0$}. The motivation for using the 
deviance version of likelihood comes from the fact that for the 
normal-like distribution probabilities the deviance is equivalent to the 
usual $\chi^2$ estimator. Nevertheless, the minimization of Poisson-like 
distribution probabilities can not be formulated in terms of a simple 
$\chi^2$ formalism.

The solution to the minimal deviance (or, in other words, maximal 
likelihood)
\begin{equation}
\mathcal{D} = \text{min}.
\end{equation}
is usually found by solving for an extreme
\begin{equation}
\nabla\mathcal{D} \equiv 0,
\end{equation}
where the gradient $\nabla$ is formed by derivatives over the whole 
parameter space. In our two-measurement model, the deviance (likelihood) 
depends on the following parameters: the $\alpha$ and $\beta$ 
coefficients from the analog-to-digital conversion $A(p)$, the variance 
of the analog signal $\gamma^2$, and the dead-time fraction $\delta$ of 
the photon counter. In addition to these four model parameters, the 
deviance depends also on all (unknown) numbers of incoming photons 
$p_i$. In general, the deviance thus has $N+4$ parameters for $2N$ data 
points (analog and photon counts).

Due to the particular structure of the deviance in Eq.~\eqref{deviance}, 
we can split the global minimization procedure for $\mathcal{D}$ in two 
parts: the outer part drives the minimization over $\alpha$, $\beta$, 
$\gamma^2$ and $\delta$ parameters, and the inner part deals with the 
``nuisance'' parameters $p_i$ for each iteration of the outer part.

Assuming that the outer part already supplies parameters $\alpha$, 
$\beta$, $\gamma^2$ and $\delta$, the inner part proceeds as follows: 
since in the total deviance $\mathcal{D}$ only the $i$th term 
$\mathcal{D}(a_i,m_i,p_i)$ depends on particular $p_i$, we can simplify 
the $p_i$ part of its gradient,
\begin{equation}
\frac{\partial\mathcal{D}}{\partial p_i} =
  \frac{\partial\mathcal{D}(a_i,m_i,p_i)}{\partial p_i},
\label{grad}
\end{equation}
by introducing a marginalized (conditional) number of photons,
\begin{equation}
\tilde{p}_i = \argmin_{p_i}\mathcal{D}(a_i,m_i,p_i),
\end{equation}
and profile deviance (as in profile likelihood)
\begin{equation}
\hat{\mathcal{D}}(a_i,m_i) =
  \min_{p_i} \mathcal{D}(a_i,m_i,p_i) =
  \mathcal{D}(a_i,m_i,\tilde{p}_i).
\label{profile}
\end{equation}
Solving this equation for all $p_i$ produces a total deviance without 
the nuisance parameters $p_i$. Finally, the deviance is contracted into
\begin{equation}
\hat{\mathcal{D}} = \sum_i\hat{\mathcal{D}}(a_i,m_i),
\label{marginalized-deviance}
\end{equation}
which depends only on the remaining four parameters $\alpha$, $\beta$, 
$\gamma^2$, and $\delta$, and is minimized by the outer part of the 
procedure.

For our two particular measurement models, 
Eqs.~\eqref{grad}--\eqref{profile} in the inner part of the minimization 
correspond to finding a suitable root of a polynomial of the fourth 
order in $p_i$.  Although analytical solutions exist, they are not very 
practical for real application. The solution can be obtained with the 
Newton-Raphson iteration
\begin{equation}
\tilde{p}_i^{[j+1]} =
  \tilde{p}_i^{[j]} -
  \frac{\mathcal{D}'(a_i,m_i,\tilde{p}_i^{[j]})}
       {\mathcal{D}''(a_i,m_i,\tilde{p}_i^{[j]})}
\label{newton}
\end{equation}
where $\mathcal{D}'$ and $\mathcal{D}''$ are respectively the first and 
the second derivative of $\mathcal{D}(a_i,m_i,p_i)$ with respect to 
$p_i$ in Eq.~\eqref{deviance1}.  The iteration in Eq.~\eqref{newton} is 
started with a suitable approximation $\tilde{p}_i^{[0]}$ and is 
repeated until $\left|\tilde{p}_i^{[j+1]}-\tilde{p}_i^{[j]}\right|$ 
becomes smaller than $\varepsilon$, with $\varepsilon$ set to some small 
number ($\sim10^{-6}$). Note that in some cases the minimum over $p_i$ 
is not the zero of the derivative in Eq.~\eqref{grad}, but can be 
instead found at the boundary, $p_i=0$, of the validity interval of the 
$p_i$ parameter.

The outer part of the minimization deals with the total deviance in 
Eq.~\eqref{marginalized-deviance} with respect to the remaining 
non-fixed parameters\footnote{The minimization in Eq.~\eqref{profile} is 
thus embedded inside the outer minimization.} and this can be carried 
out with a variety of nonlinear minimization procedures (see for example 
\cite{minuit}).  Denoting the final values of the parameters in the 
deviance minimum with $\tilde{\alpha}$, $\tilde{\beta}$, and 
$\tilde{\delta}$, the set of final values of nuisance parameters
\begin{equation}
\breve{p}_i =
  \tilde{p}_i(\tilde{\alpha},\tilde{\beta},\gamma^2,\tilde{\delta})
\label{most-likely}
\end{equation}
in the global minimum of the deviance represents the ultimate (most 
likely) synthesis of the analog and photon-counting modes of the lidar 
data acquisition. Note that $\gamma^2$ is kept fixed at the value of the 
initial approximation throughout this procedure.

\begin{figure}[t]
\centering
\includegraphics[width=\linewidth]{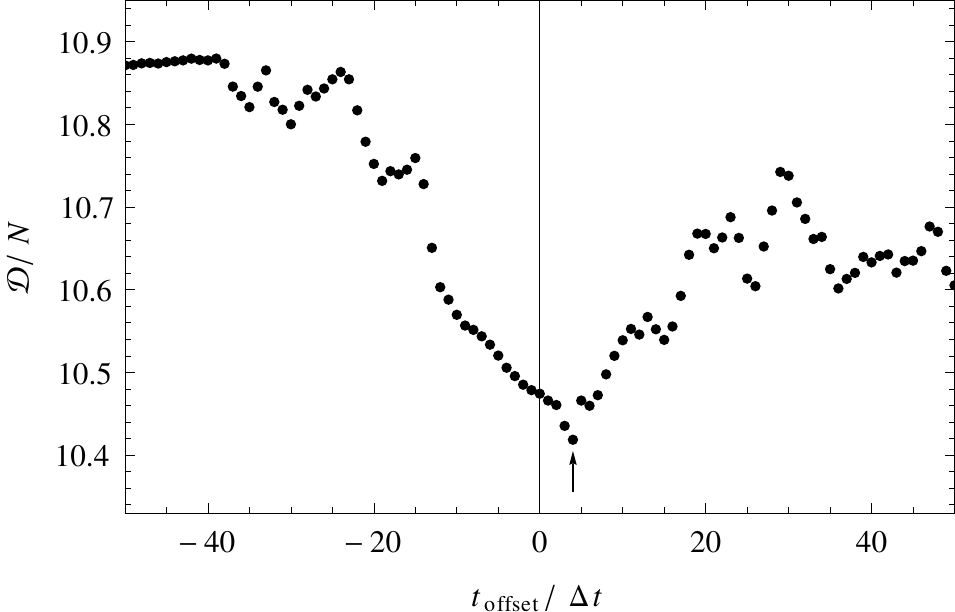}
\caption{Dependence of the normalized deviance $\mathcal{D}/N$ on 
relative offset $t_\text{offset}$ between the analog and photon-counting 
traces. The arrow indicates the position of the minimum at 
$t_\text{offset}=4\Delta t=\unit[100]{ns}$.}
\label{f:offset}
\end{figure}

\subsection{Relative acquisition delay}

\begin{figure}[t]
\centering
\includegraphics[width=\linewidth]{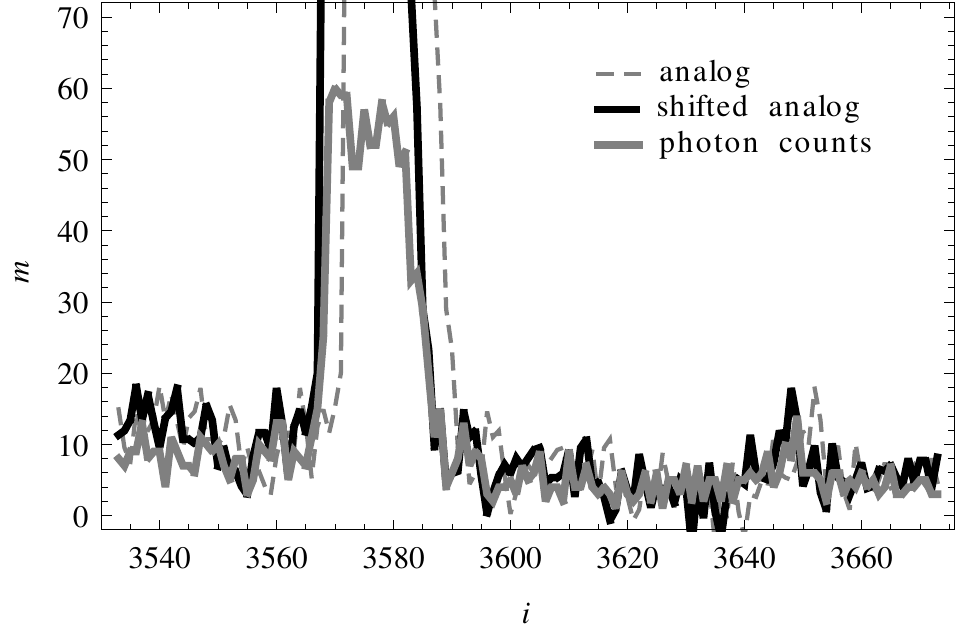}
\caption{Haze feature before and after the shift of the analog trace.  
The interval $[3450, 3660]$ in $i$ corresponds to $[13.275, 13.725]$\,km 
in range.}
\label{f:offset-feature}
\end{figure}

In the acquisition system the input signal flows through quite different
electronic sub-components of the transient recorders (e.g.\ see schematic 
in manual \cite{licel}) so expecting hardware and firmware related 
differences in delay time is highly justified. The final analog and 
photon counting traces can thus be subject to substantial relative 
offset. In the case of our particular recorder this offset amounts to 
four sample times, $t_\text{offset}=4\,\Delta t=\unit[100]{ns}$. The 
dependence of the total deviance on $t_\text{offset}$ is shown in 
Fig.~\ref{f:offset}.  How this offset is influencing the details of the 
example trace can be observed in Fig.~\ref{f:offset-feature}. The two 
haze features found around \unit[13.5]{km} in the analog and 
(uncorrected) photon-count modes perfectly match after the 
$t_\text{offset}$ shift.  The same holds for the small noise-like 
features in the rest of the trace, mostly responsible for the distinct 
minimum of deviance in Fig.~\ref{f:offset}.

\subsection{Parameter bias due to data distribution}

Typical lidar returns do not cover uniformly the whole dynamic range 
available in analog, $a$, and photon-counting, $m$, modes. Most of the 
data resides in the tail of the lidar return, occupying only the 
lower-left sector of the $(a,m)$ phase space (c.f.\ Fig.~\ref{f:am}).  
The large contribution of this sector can thus introduce a bias in the 
likelihood maximization procedure. Furthermore, the acquisition 
parameters are influenced by the different parts of the $(a,m)$ phase 
space. Analog baseline $\beta$ is well defined by the lidar tail (lower 
left sector in Fig.~\ref{f:am}), the dead-time fraction $\delta$ is 
mostly sensitive to the large-signal parts (upper right sector), and the 
photon-to-analog coefficient $\alpha$ is mostly influenced by the small 
and medium part of the trace (left side in Fig.~\ref{f:am}). To remove 
and quantify this bias we can bin the data with several different 
partitions of the $(a,m)$ phase space and balance the relative 
contribution of each data point $(a_i,m_i)$ to the total deviance with a 
weight $w(a_i,m_i)=w_j$, where $j$ is the appropriate bin index. To 
maintain the correspondence with the previous non-binned case 
(equivalent to the binning with $w_j=1$) all binning variants are 
required to satisfy a summation rule $\sum_j w_j=N$, with $N$ being the 
number of data points.  In all binning variants the weight should be 
proportional to the inverse density so that all points in a particular 
bin contribute to the deviance with the same weight as the other 
non-empty bins. With these two requirements the weights are obtained as
\begin{equation}
w(a_i,m_i) = w_j = \frac{N}{N_\emptyset N_j},
\label{weigths}
\end{equation}
where $N_\emptyset$ is the number of non-empty bins and $N_j$ the number 
of data points in a particular bin $j$.

\begin{figure}[t]
\centering
\includegraphics[width=0.9\linewidth]{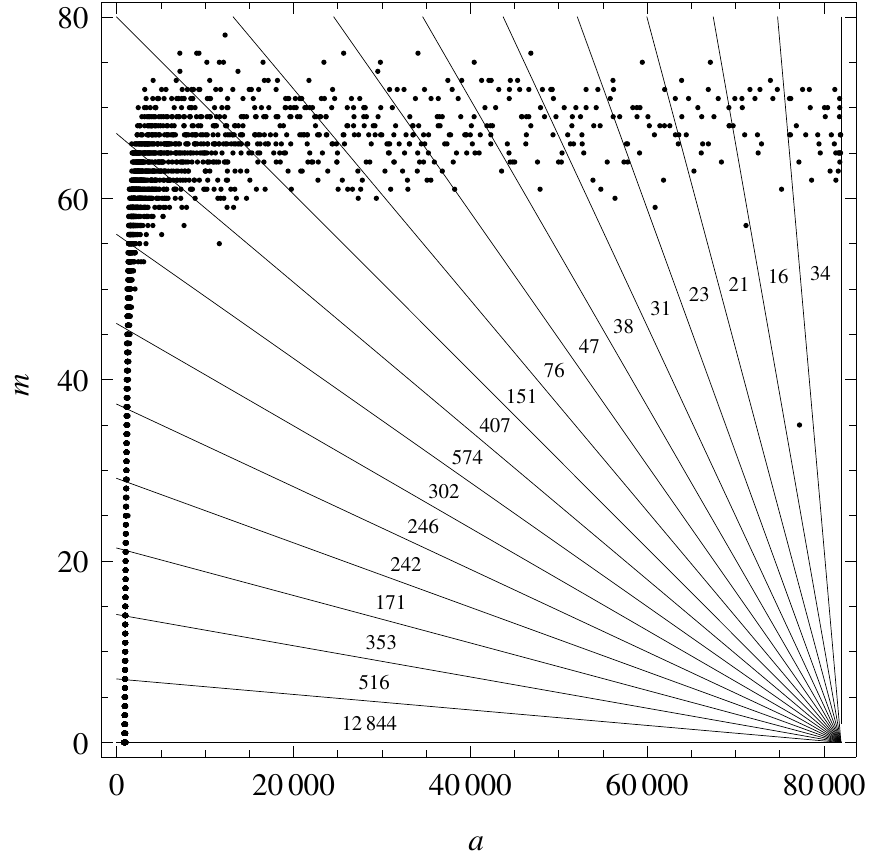}
\caption{Example of a debiasing binning of data points $(a_i,m_i)$ from 
the example lidar return into the fan-like bins. Note that the bottom 
bin contains many more data points than all of the other bins.  The 
total deviance weights are set proportionally to the inverse of the 
particular bin count $N_j$, which is also shown for all bins.}
\label{f:debias}
\end{figure}

The debiased version of deviance from Eq.~\eqref{marginalized-deviance} 
is now written as
\begin{equation}
\hat{\mathcal{D}} =
  \sum_i w(a_i,m_i)\,\hat{\mathcal{D}}(a_i,m_i).
\label{debiased-deviance}
\end{equation}

\begin{figure}[t]
\centering
\includegraphics[width=0.9\linewidth]{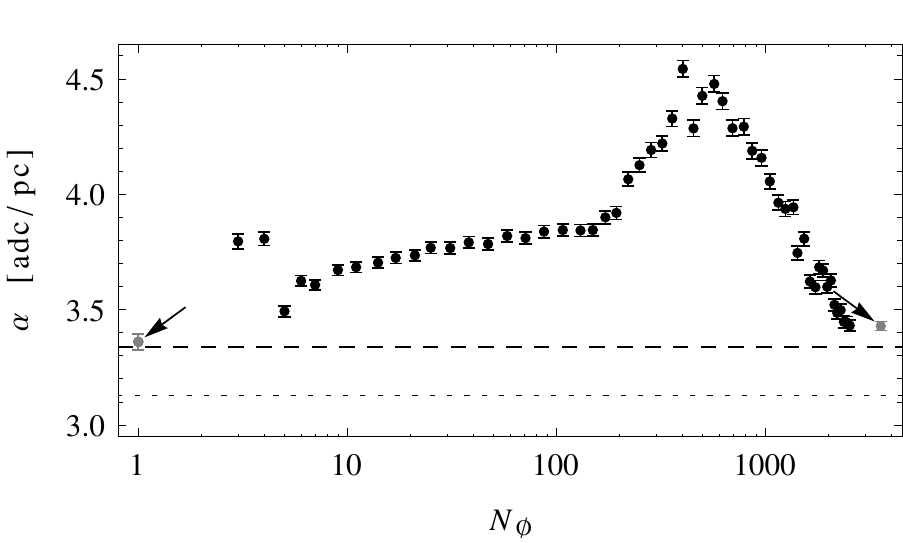}
\\
\includegraphics[width=0.9\linewidth]{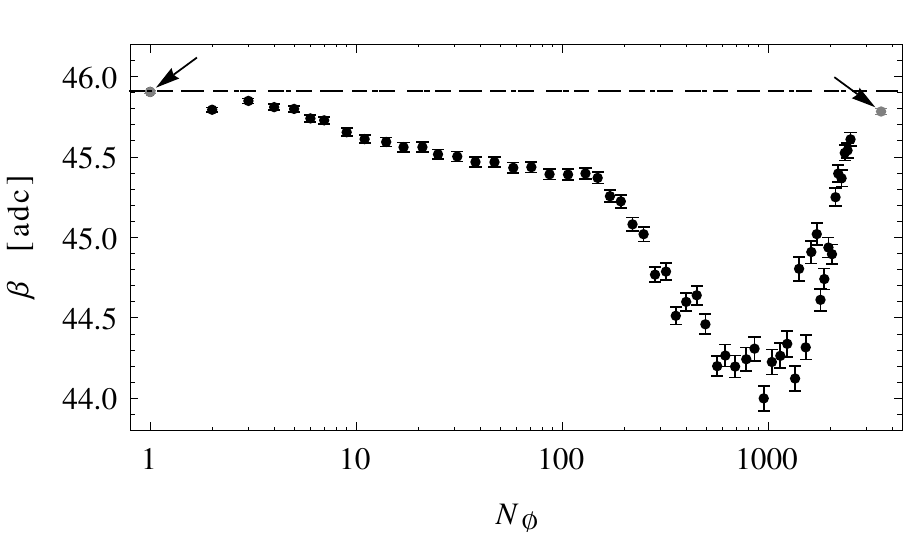}
\\
\includegraphics[width=0.9\linewidth]{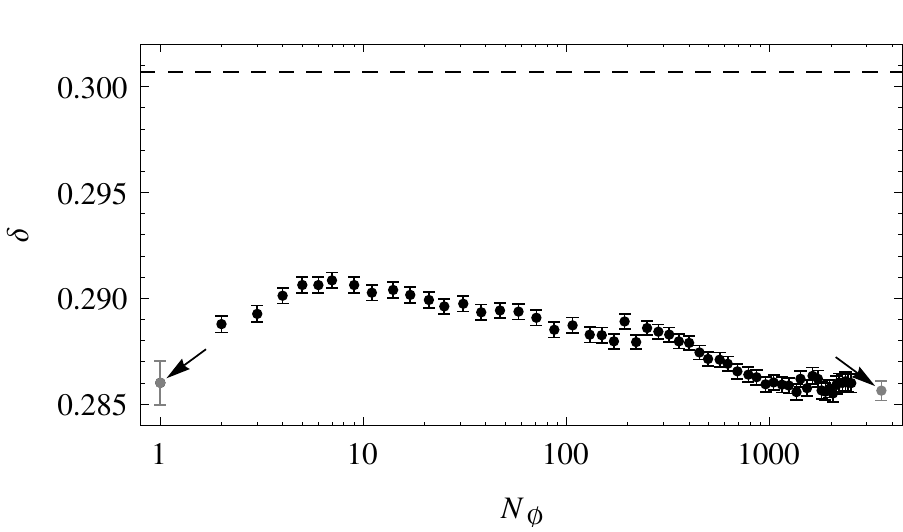}
\caption{Dependence of the analog parameters $\alpha$, $\beta$ and
photon-count parameter $\delta$ on the number of non-empty bins
$N_\emptyset$ for the fan-like binning (black points). Left and right 
arrows indicate parameter values from the un-binned and the 
infinitely-fine binned cases, respectively. The values from the initial 
estimates are shown in dashed and dotted lines (see the text for more 
details).}
\label{f:binning-dependence}
\end{figure}

Here we consider several variants of the binning divisions.
\begin{itemize}
\item Un-binned case, $w(a_i,m_i)=1$: every data point is considered 
with the same weight. Note that in this way in cases of lidar traces 
with long tails, the overwhelming contribution will come from the 
points with small values in both modes $a_i$ and $m_i$.
\item Fan-like binning: data is categorized into a histogram with 
fan-like bin shapes radiating from the lower right corner, 
$(a_\text{max}, m_\text{min})\equiv(2^{12}, 0)$ (see 
Fig.~\ref{f:debias}--left).
\item Infinitely-fine binning: since both measurement modes $a_i$ and 
$m_i$ have only discrete integer values, we treat each of these discrete 
points with the same weight.
\end{itemize}

In the top two panels of Fig.~\ref{f:binning-dependence} initial 
estimates for the analog parameters $\alpha$ and $\beta$ are shown for 
two versions of the $\chi^2$ fit from Eq.~\eqref{chi2}. The dashed lines 
are for the $\chi^2$ with squares of $a_i-A(m_i)$ (involving uncorrected 
photon-counts) and the dotted lines are for a version where the 
dead-time fraction $\delta$ is estimated first and then the $\chi^2$ is 
constructed with dead-time corrected photon counting, 
$a_i-A(C^{-1}(m_i))$. In the bottom panel of 
Fig.~\ref{f:binning-dependence} the initial estimate for dead time 
$\delta$ is shown as a dashed line. The arrows on the left and the right 
of all panels are indicating the results of the un-binned and the 
infinitely-fine binned cases, respectively. The rest of the points 
correspond to the fan-like binning for different granularity of the 
binning represented by the number of non-empty bins $N_\emptyset$.

While there are only minor differences between the un-binned and 
infinitely binned results, the fan-like binning exhibits large 
variations of the order of 25\% for $\alpha$, 5\% for $\beta$, and 15\% 
for $\delta$ when changing the binning size\footnote{Uncertainties are 
in fact not so large, considering that the parameters are obtained on a 
\emph{single} trace with $N_\text{s}=20$ summation only.}.  
Nevertheless, with increasingly fine binning the parameters tend to 
converge to the two values obtained by the un-binned and infinitely 
binned cases, indicating that (at least for this trace depth) the 
maximum likelihood method is only mildly biased by the particular data 
distribution. On the other hand, in longer traces the distribution of 
the data in the lower left sector might influence the reconstruction, 
especially if the measurement models are not accurate enough.

\section{Results}

\subsection{Reconstructed number of photons}

In order to follow the evolution of the reconstructed (most likely) 
number of photons $\breve{p}$ from Eq.~\eqref{most-likely} let us 
introduce a transition indicator
\begin{equation}
u = \frac{p_m - \breve{p}}{p_m - p_a},
\label{transition}
\end{equation}
where $p_a$ and $p_m$ are direct estimates for photons obtained from the 
two measurements,
\begin{equation}
p_a = A^{-1}(a) = \frac{a-\beta}{\alpha}
\end{equation}
is the analog-to-photon conversion (inverse of Eq.~\eqref{analog}) and
\begin{equation}
p_m = C^{-1}(m) = \frac{m}{1-\delta m}
\end{equation}
is the corrected photon count (inverse of Eq.~\eqref{dtc-mean}). Note 
that the latter works only for $m<1/\delta$. The values of the indicator 
$u$ defined in this way will be close to 1 when the reconstructed number 
of photons $\breve{p}$ is close to the prediction from the current 
inversion $p_a$. On the other hand, $u$ will be close to 0 when 
$\breve{p}$ is close to the dead-time corrected photon-count $p_m$.  
Values between 0 and 1 are indicating transition between the two 
measurements.

\begin{figure}[t]
\centering
\includegraphics[width=\linewidth]{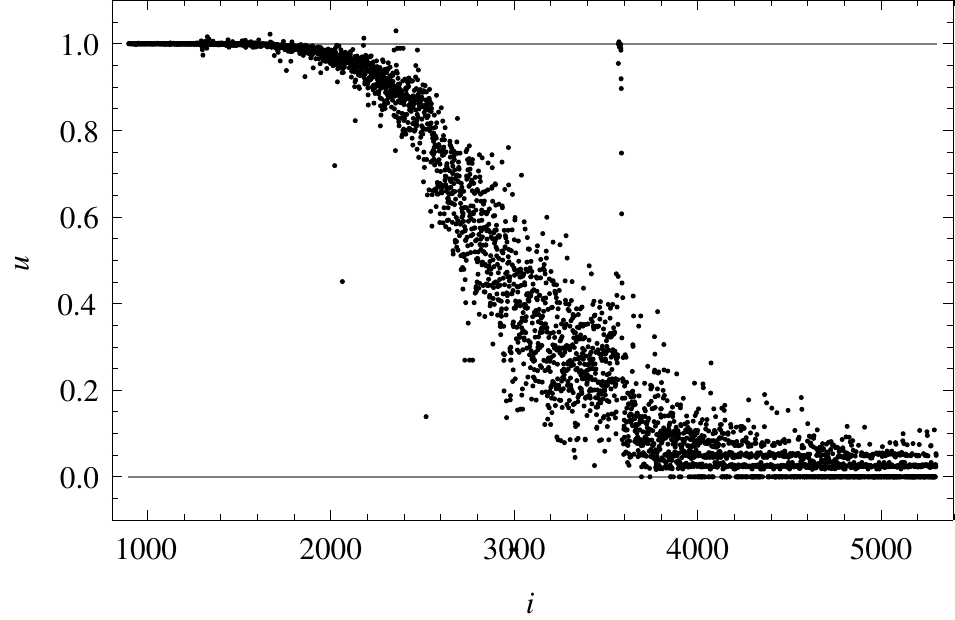}
\caption{Behavior of the transition indicator $u$ from
Eq.~\eqref{transition}.  The $[1500, 5000]$ interval in $i$ corresponds
to the $[5.625, 18.75]$\,km interval in range.}
\label{f:transition}
\end{figure}

In Fig.~\ref{f:transition} the changing of the value of indicator $u$ 
along our example trace is shown. We can clearly identify three regions 
of the indicator's behavior.

For indices $i$ below $\sim2000$ the indicator $u\approx1$ and thus the 
reconstructed number of photons is closely following the estimate from 
the analog signal. In this region the photon counting is saturated and, 
due to the dead-time, its resolution is heavily suppressed. The maximum 
likelihood method thus shifts the result towards the more accurate 
measurement: the analog signal.

For indices $i$ above $\sim4000$ the indicator $u\approx0$ and the 
reconstructed number of photons is closely following the 
photon-counting. Here the analog signal is diving into the noisy region 
around the analog baseline where SNR is small. On the other hand, the 
photon counting is far from being effected by the dead-time and thus 
maximum likelihood gives it deserved emphasis.

In the intermediate region, for indices $i$ between $\sim2000$ and 
$\sim4000$ where $0<u<1$, the reconstructed number of photons lies 
between the two measurements which both have degraded accuracy, analog 
signal due to poor SNR and photon-counting due to the dead-time effects.  
Nevertheless, the value of $\breve{p}$ is chosen according to the 
maximum likelihood, effectively combining the two less accurate 
measurements into one with smaller error\footnote{e.g.\ 
maximum-likelihood combination of two normally distributed measurements 
with errors $\sigma_1$ and $\sigma_2$ gives a new estimate with a 
smaller error 
$\sigma_1\sigma_2/\sqrt{\smash[b]{\sigma_1^2+\sigma_2^2}}$}.

Note that all variants of the standard ''gluing'' procedures would in 
this picture produce a step-like functional form of indicator $u$, 
abruptly crossing over from 1 to 0 at a position that depends on a 
particular choice of the ``gluing'' method.

\subsection{Multiple lidar traces and dynamic range}

Standard processing of the lidar returns usually employs extensive 
stacking (summation) of the lidar traces in order to increase the SNR 
ratio. Since the maximum likelihood method has no built-in notion of 
range and data-point ordering, multiple traces can instead be 
concatenated in order to increase the accuracy of the reconstruction.  
All points from different traces can thus be treated equally and 
processed at the same time, as long as the acquisition parameters 
$\alpha$, $\beta$, $\gamma^2$, and $\delta$ are considered stable in the 
respective time frame of the recording of the traces.  Nevertheless, 
this procedure will suffer a slowdown linear with the number of data 
points used\footnote{At the time of the writing, it takes \unit[0.2]{s} 
per 16k trace on a normal desktop computer.}.

In case of our equipment, relative scatter of the reconstructed 
parameters within individual \unit[480]{s} runs were below 1.6\% for 
$\alpha$, 0.24\% for $\beta$, and 0.28\% for $\delta$. Relative 
differences between the mean values of the reconstructed parameters for 
the runs from the beginning and the end of the measurement campaign were 
below 0.61\% for $\alpha$, 0.54\% for $\beta$, 1.62\% for $\gamma^2$, 
and 0.18\% for $\delta$. In such stable conditions it is thus possible 
to concatenate a large number of lidar traces in order to increase the 
accuracy of the reconstructed photon returns.

In the usual lidar operation, the gain of the analog channel is set to a 
level which produces a discretized trace with the maximum signal 
slightly below (or, like in our case, above) the ADC saturation limit.  
In this way the analog signal can cover the whole dynamic range of the 
ADC, exposing nicely the saturation of the photon counting (see 
Fig.~\ref{f:am}). It turns out that if for whatever reason the analog 
signal is covering only a smaller fraction of the available dynamic 
range or that the amount of photons is not saturating the photon 
counting, the procedure described above will still produce stable and 
reasonable estimates of the $\alpha$, $\beta$, and $\gamma^2$ parameters 
since they are anyway dominated by the data in the lower-left corner of 
the $(a_i,m_i)$ diagram. On the other hand, the dead-time fraction 
$\delta$ will in this case tend to zero (towards the ideal counter),
since the absence of the photon-counting saturation in the data 
effectively gives an estimate $1/\delta\to\infty$, but all this without 
actually influencing the reconstructed number of photons $p_i$.

\section{Discussion and conclusions}

Modern transient recorders offer two principally different measurements 
of the same photon return: the digitization of the analog signal and the 
photon-counting mode. We are thus challenged, not only to use them in 
their respective regions of validity (like the usual ``gluing'' methods) 
but to combine them into a more accurate estimate of the photon numbers 
by using detailed statistical models of both acquisition processes.

The maximum likelihood procedure described in this work offers a
reconstruction of photon returns that has a natural transition between 
the analog and photon-counting signals and is based on their analytical 
measurement models. In this work we have been using fairly rudimentary 
models of the two measurement processes, nevertheless they still 
adequately capture the main strengths of the new reconstruction 
procedure. Furthermore, if more detailed models are needed, they can be 
simply included into the probability expressions entering the likelihood 
function.

In this method we are strongly discouraged to attempt any kind of 
background subtraction. The background photons are treated as a normal 
signal since they appear in both measurement modes as viable data. Any 
removal of background (from dawn or daylight) should be done on the 
final reconstructed photon numbers.

The maximum likelihood method works in all conditions, even in the 
presence of clouds or other enhanced scattering objects. It fails only 
when the input levels of photons are not exploring the whole dynamic 
range of the transient recorder (i.e.\ both corners of Fig.~\ref{f:am}) 
and thus no reliable estimate of the acquisition parameters $\alpha$, 
$\beta$, $\gamma^2$, and $\delta$ can be obtained. Through the offset 
analysis of the likelihood value it offers a simple way for estimation 
of the potential relative delay between the two measurement traces, 
which in our particular case turned out not to be negligible.

The code implementing the maximum-likelihood reconstruction of lidar 
returns is available under the GPL3 license at 
\href{http://www.ung.si/~darko/lidar/}{http:/\!\!/www.ung.si/\textasciitilde 
darko/lidar/}.

\section*{Acknowledgments}

Author wishes to thank Matej Horvat and Martin O'Loughlin for fruitful 
discussions and Fei Gao from the Center for Atmospheric Research of 
University of Nova Gorica for recording the actual lidar return used in 
the examples.  The research was supported by the Ministry for Higher 
Education, Science, and Technology of Slovenia and the Slovenian 
Research Agency.

\appendix

\section{Dead-time counter}
\label{a:dtc}

For an ideal counter with sampling time $\Delta t$ the discrete 
probability distribution of the number of counts $k$ for a Poisson 
process with rate $r$ and mean $M=r\Delta t$ is given by 
$\mathcal{P}_k(M)$ where
\begin{equation}
\mathcal{P}_k(x) = \frac{x^k e^{-x}}{k!}.
\label{poisson}
\end{equation}
For a counter with non-extending dead-time $\tau$ \cite{axton} the 
corresponding probability distribution can be found in 
Ref.~\cite{mueller}.  Restructuring the equations and performing partial 
summations of the expressions given there, the probability $W_k$ of 
observing $k$ counts now becomes
\begin{equation}
W_k =
  \frac{1}{1+M\delta}\left[
    R_{k-1} - 2R_k + R_{k+1} + \Delta_k
  \right],
\label{dtc-pdf}
\end{equation}
where $\delta=\tau/\Delta t$, the short-hand $R_k=R_k(t_k)$ and the 
truncated mean for $k$ counts is given by
\begin{equation}
t_k = M(1-k\delta).
\end{equation}
$R_k=R_k(t_k)$ is fully expressed as
\begin{align}
R_k(x) &=
  U(x) \sum_{j=0}^{k-1} (k-j) \mathcal{P}_j(x) =
\nonumber
\\
  &= U(x)[(k-x)\,Q(k,x) + k\mathcal{P}_k(x)],
\end{align}
where $Q(j,x)=\Gamma(j,x)/\Gamma(j)$ is the \emph{regularized} upper 
incomplete Gamma function \cite{walck}, with the upper incomplete Gamma 
function
\begin{equation}
\Gamma(a,x) = \int_x^\infty u^{a-1}e^{-u}\,{\rm d}u
\end{equation}
and $\Gamma(a) = \Gamma(a,0)$.
The unitary step function $U(x)$ is defined in the usual way,
\begin{equation}
U(x) =
\begin{cases}
1 & \text{if $x>0$},
\\
0 & \text{otherwise},
\end{cases}
\end{equation}
and the remainder $\Delta_k$ in Eq.~\eqref{dtc-pdf} is given explicitly 
as
\begin{equation}
\Delta_k =
\begin{cases}
0 & \text{if $k\leqslant K-1$},
\\
(K+1)(1+M\delta)-M & \text{if $k=K$},
\\
M-K(1+M\delta) & \text{if $k=K+1$}.
\end{cases}
\end{equation} 

The upper limit on possible counts depends on the dead-time fraction,
\begin{equation}
K = \lfloor1/\delta\rfloor,
\end{equation}
where $\lfloor x\rfloor$ is denoting the largest integer smaller (and 
not equal) than $x$.

The mean dead-time count $m$ can be expressed in terms of the ideal 
count $M$ as
\begin{equation}
m = C(M) = \frac{M}{1+M\delta}.
\end{equation}
The exact expression for the variance of the dead-time counter is
\begin{align}
V_\delta =&
  \frac{2}{1+M\delta}
  \sum_{k=0}^{K} \left[(k-t_k)\,Q(k,t_k) + k\mathcal{P}_k(t_k)\right] +
\nonumber
\\
  &+ H(m-K),
\label{exact}
\end{align}
where the ``hump'' function is defined as $H(x)=x(1-x)$.

Using $d=M\delta$ as a mean ``lost'' count, for $d\ll1$ the exact 
variance from Eq.~\eqref{exact} asymptotically \cite{mueller} behaves as
\begin{equation}
V_\delta \approx
  M\left[
  \frac{1}{(1 + d)^3} +
  \frac{\mu^2(6 + 4d + d^2)}
       {6M(1 + d)^4}
  \right],
\label{asymptotic}
\end{equation}
where the part in the square brackets is the suppression factor relative 
to the Poisson variance $V_\text{Poisson}=M$. For $d\gg1$ the variance 
is well described by a fully saturated dead-time counter,
\begin{equation}
V_\delta \approx
  H(m-K) =
  H(\operatorname{frac}(m)),
\label{saturated}
\end{equation}
where $\operatorname{frac}(x)=x-\lfloor x\rfloor$ is the function 
returning fractional (non-integer) part of an argument. Note that the 
expression for asymptotic variance in Eq.~\eqref{asymptotic} converges 
in this regime to 1/6 and thus well describes the average of the 
oscillatory dependence of $V_\delta$ on $\delta$ in Eqs.~\eqref{exact} 
and \eqref{saturated}.


\begin{thebibliography}{99}

\bibitem{eichinger} V.~A.~Kovalev and W.~E.~Eichinger, \emph{Elastic 
Lidar} (Wiley, 2004), pp.~136--141.

\bibitem{donovan} D.~P.~Donovan, J.~A.~Whiteway, and A.~I.~Carswell,
``Correction for nonlinear photon-counting effects in lidar systems'',
Appl.\ Opt.\ {\bf 32}, 6742--6753 (1993).

\bibitem{gluing} Z.~Liu, Z.~Li, B.~Liu, and R.~Li, ``Analysis of 
saturation signal correction of the troposphere lidar'', Chin.\ Opt.\ 
Lett.\ {\bf 7}, 1051--1054 (2009).

\bibitem{newsom} R.~K.~Newsom, D.~D.~Turner, B.~Mielke, M.~Clayton, 
R.~Ferrare, and C.~Sivaraman, ``Simultaneous analog and photon counting 
detection for Raman lidar'', Appl.\ Opt.\ {\bf 48}, 3903--3914 (2009).

\bibitem{whiteman} D.~N.~Whiteman, B.~Demoz, P.~Di~Girolamo, J.~Comer,
I.~Veselovskii, K.~Evans, Z.~Wang, M.~Cadirola, K.~Rush, G.Schwemmer, 
B.~Gentry, S.~H.~Melfi, B.~Mielke, D.~Venable, and T.~Van~Hove, ``Raman
lidar measurements during the international H$_2$O project. Part I: 
Instrumentation and analysis techniques'', J.\ Atmos.\ Oceanic Technol.\
{\bf 23}, 157--169 (2006).

\bibitem{licel-gluing} B.~Mielke, ``Analog + photon counting'',
\\
\href{http://www.licel.com/analogpc.pdf}{http:/\!\!/www.licel.com/analogpc.pdf}.

\bibitem{licel}
\href{http://www.licel.com/Transientrecorder.pdf}{http:/\!\!/www.licel.com/Transientrecorder.pdf};
\href{http://www.licel.com/TRInstallation.pdf}{http:/\!\!/www.licel.com/TRInstallation.pdf}.

\bibitem{minuit} F.~James, ``Minuit, Function Minimization and Error
Analysis'', CERN long writeup D506 (1998); and implementation in
\href{http://root.cern.ch}{http:/\!\!/root.cern.ch}.

\bibitem{axton} E.~J.~Axton and T.~B.~Ryves, ``Dead-time corrections in
the measurement of short-lived radionuclides'', Int.\ J.\ Appl.\
Radiation Isotopes {\bf 14}, 159--161 (1963).

\bibitem{mueller} J.~W.~M\"uller, ``Some formulae for a 
dead-time-distorted Poisson process'', Nucl.\ Instr.\ Methods {\bf 117}, 
401--404 (1974).

\bibitem{walck} C.~Walck, ``Hand-book on statistical distributions for 
experimentalists'', Stockholms Universitet, Internal Report 
SUF-PFY/96-01, 10 September 2007, pp.~159--160.

\end{thebibliography}
\end{document}